# Exploring the spatiotemporal heterogeneity in the relationship between human mobility and COVID-19 prevalence using dynamic time warping

Hoeyun Kwon[1], Kaitlyn Hom[2], Mark Rifkin[3], Beichen Tian, Caglar Koylu[1]

[1]Department of Geographical and Sustainability Sciences, University of Iowa, Iowa City, IA, USA
[2]Dougherty Valley High School, San Ramon, CA, USA
[3]Francis Parker School, San Diego, CA, USA
Email: {hoeyun-kwon, kaitlyn-hom, mark-rifkin, beichen-tian, caglar-koylu}@uiowa.edu
https://www.geo-social.com

## Abstract

Understanding where and when human mobility is associated with disease infection is crucial for implementing location-based health care policy and interventions. Previous studies on COVID-19 have revealed the correlation between human mobility and COVID-19 cases. However, the spatiotemporal heterogeneity of such correlation is not yet fully understood. In this study, we aim to identify the spatiotemporal heterogeneities in the relationship between human mobility flows and COVID-19 cases in U.S. counties. Using anonymous mobile device location data, we compute an aggregate measure of mobility that includes flows within and into each county. We then compare the trends in human mobility and COVID-19 cases of each county using dynamic time warping (DTW). DTW results highlight the time periods and locations (counties) where mobility may have influenced disease transmission. Also, the correlation between human mobility and infections varies substantially across geographic space and time in terms of relationship, strength, and similarity.

## 1. Introduction

As COVID-19 transmits through exposure to respiratory fluids from infected people, control measures to delay the spread of COVID-19 have been implemented to limit people's mobility with travel restrictions and nonessential business closures. The effectiveness of such policies relies heavily on our understanding of the relationship between human mobility and disease transmission. Using human mobility changes derived from anonymous mobile phone location data, previous studies have found a positive correlation between COVID-19 infections and human mobility (Kraemer *et al*. 2020; Miller *et al*. 2020; Schlosser *et al*. 2020). These studies adopted a variety of mobility indicators such as the distance traveled by mobile phone users, stay duration in an area, and entropy of activities (Xu *et al.* 2021). These mobility indicators are useful to extract a global time-varying relationship between mobility and disease outcomes. For example, Xiong *et al*. (2020)'s time-series modeling revealed a global relationship between mobility and disease cases in relation to lock-down and travel restrictions at the national level. However, little is known about the spatiotemporal non-stationarity or heterogeneity in such a relationship, in other words, where and when human mobility may be associated with disease infections. Such information may be crucial for implementing location-based health care policies and interventions (Thomas *et al.* 2020).

In this study, we aim to identify the spatially heterogeneous relationships between human mobility and COVID-19 infections in U.S. counties. We do this by comparing the time-series of COVID-19 daily cases for each county to human mobility flows within and into each county using dynamic time warping (DTW) (Brown and Rabiner 1982). First, we extract COVID-19 disease cases for counties during the largest surge between November 23, 2020 and January 24, 2021 (9 weeks). Second, we compute a mobility measure that aggregates flows within and into each county using SafeGraph mobile phone location data (SafeGraph 2020). Finally, we perform DTW on the 7-day average of daily disease and mobility time series and examine the spatial heterogeneity by visualizing the DTW results.

## 2. Related Work

Mobile phone data have been effectively used to extract human mobility flows at different scales (i.e., census block, county, state) during the COVID-19 pandemic (Kang *et al.* 2020). Lockdown orders have been implemented in the U.S. since mid-March, which resulted in a substantial decrease in human mobility. Specifically, a significant decrease in travel time (Borkowski *et al.* 2021) and a long-term reduction of long-distance mobility were found to be caused by lockdowns (Schlosser *et al.* 2020). Findings of these studies suggested that lockdowns limited human mobility, and thus, may have helped contain the spread of COVID-19. On another note, studies conducted at the global scale revealed that mobility and transmission rates were found to be positively correlated across 52 countries (Nouvellet *et al.* 2021). In addition to revealing a positive association between mobility and COVID-19 infections at global or national scales, other studies revealed that spatial heterogeneity of populations at local scales produced substantial variations in COVID-19 infection severity and peak infection times (Thomas *et al.* 2020). Similarly, a case study on two counties found that COVID-19 reproduction rates varied greatly due to age and racial heterogeneity within regions of the county (Hou *et al.* 2021). These studies suggest that spatial heterogeneity has a substantial effect on COVID-19 case rates and spread, especially in highly populated metropolitan areas.

## 3. Methodology

### 3.1 Data processing

We use COVID-19 cases and human mobility flow of each county in the contiguous U.S. Daily number of COVID-19 new cases using a 7-day rolling average of each county was obtained from U.S. COVID Risk & Vaccine Tracker (Act Now Coalition 2020). Human mobility data were derived from SafeGraph mobile phone location data between counties in the U.S. (SafeGraph 2020). Our study period is from November 30, 2020, to January 24, 2021, which covers the highest peak of COVID-19 cases in the U.S. As it is examined that there is a 7-day lag between mobility inflow and COVID-19 case increase (Xiong *et al*. 2020), we also apply a 7-day lag between human mobility and COVID-19 cases to capture the potential COVID-19 transmissions due to human contacts. To avoid biased results caused by small counties with both low mobility and a small number of COVID-19 cases, we only include 1,175 counties in metropolitan statistical areas delineated by the U.S. Office of Management and Budget. We then filter out 593 counties whose average number of COVID-19 cases during our study period (i.e., 56 cases) is less than the median value of the cases in all counties.

### 3.2 Human mobility flow estimation

We apply equation 1 to estimate the human mobility between an origin county O and a destination county D (Kang *et al*., 2020):

$$Mobility\ (O, D) =\ Devices\ (O, D) * Population(O)/Devices\ (O) \qquad (1)$$

Mobility from O to D is estimated by multiplying the number of devices that moved from O to D with the population of O divided by the number of devices in O. This formula provides a bulk estimate of human mobility based on the assumption that the rest of the population that was not captured in the mobile service providers for SafeGraph has the same mobility behaviors. To highlight the movements not only within a county but also into that county, we use the mobility measure as the sum of mobility within and into each county. We perform a 7-day moving average window smoothing for the mobility measure (i.e., the sum of inflow and within flow) for each county for each day to improve the estimation for human mobility and match the temporal granularity of the 7-day average COVID-19 case counts.

### 3.3 Similarity measure in time series using Dynamic Time Warping

The similarity between the trends of human mobility and COVID-19 cases is measured using DTW. DTW is a flexible model that compares two or more time series and identifies non-linear relations between them by handling different lengths, noise, shifts, and amplitude changes (Brown and Rabiner 1982; Stübinger and Schneider 2020). DTW distance between two series becomes smaller as those two series have more similar trends. Before computing DTW distances, we normalize all series using the min-max approach and rescale them to have a fixed range of [0, 1]. Normalization allows comparing DTW values from different pairs of time-series data that may have a different range of values and units. We finally calculate a DTW distance between human mobility and COVID-19 cases for each county to investigate the similarity between the two series.

## 4. Results

To examine whether DTW distances are affected by the population size of each county, we perform a Pearson correlation analysis using population and population density (population per square mile). The relation between DTW distance and population density has the highest coefficient of 0.231, which indicates that neither population nor population density may influence the similarity between mobility and COVID-19 cases. Figure 1 highlights significant geographic variation in DTW distances. The metropolitan areas such as Los Angeles, New York City, Austin, Atlanta, Charleston, and Duluth (Minnesota) have low degrees of similarity between trends in human mobility and COVID-19 infections. Meanwhile, other metropolitan areas including Chicago, Milwaukee, St. Louis, Kansas City, Phoenix, Portland, Baltimore, and Washington D.C. have high degrees of similarity.

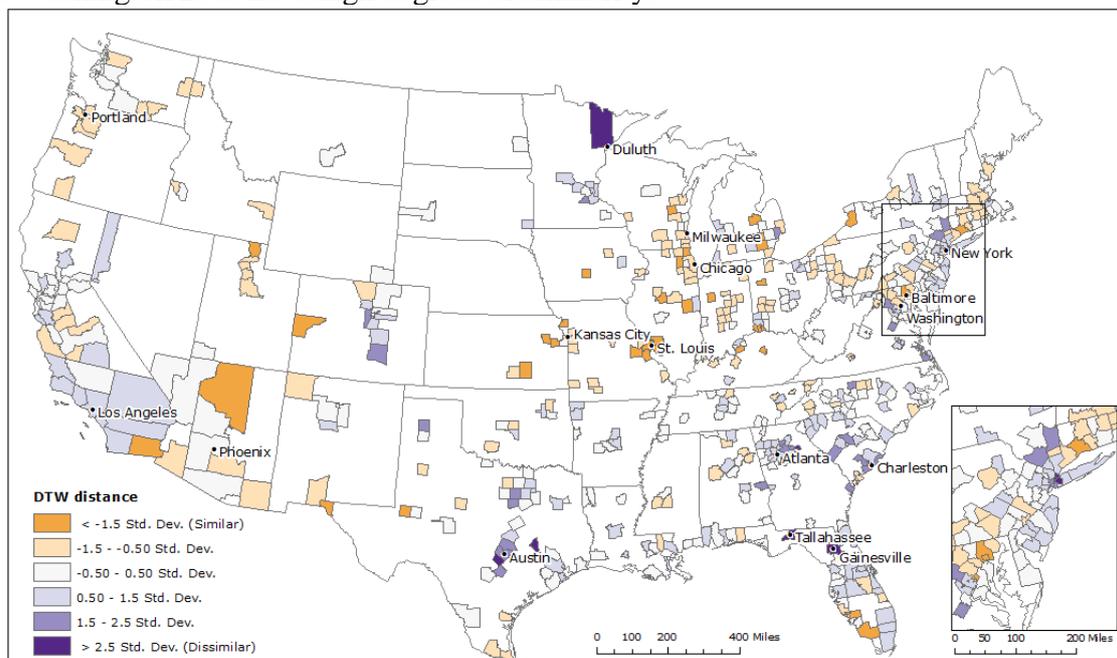

**Figure 1. DTW distances for mobility and COVID-19 infections in metropolitan counties. Negative standard deviations of values (orange) indicate similar trends of mobility and infections, while positive standard deviations (purple) indicate dissimilar trends.**

Figure 2 illustrates the time series of human mobility and COVID-19 cases of four distinct areas: Chicago, Washington, D.C., New York City, and Austin. While Chicago and Washington have similar trends in mobility and COVID-19 cases, New York City and Austin have dissimilar trends. In both areas, human mobility tends to increase during the holiday season including Christmas and New Year's Day. The graphs show clear patterns that the trend of the COVID-19 case follows the trend of mobility flow with some time lag. In contrast, COVID-19 cases in New York City keep increasing irrespective of the decreasing trend of mobility. Particularly, human mobility is decreasing suddenly right after New Year's Day which might be caused by the first case of the U.K. coronavirus variant found in New York State on January 4, 2021. However, even this sharp decrease does not suggest a potential influence on COVID-19 cases. Austin, Texas also has dissimilar trends in mobility and COVID-19 infection. Distinctively, the trends in Austin show an inverse correlation, unlike the other cities. In December, for example, human mobility decreases as COVID-19 cases go up. Since people tend to travel less when COVID-19 cases nearby increase (Gao *et al*. 2020), there can be inverse correlations between mobility and COVID-19 infections. This indicates the effect of COVID-19 on mobility behaviors, which DTW fails to detect.

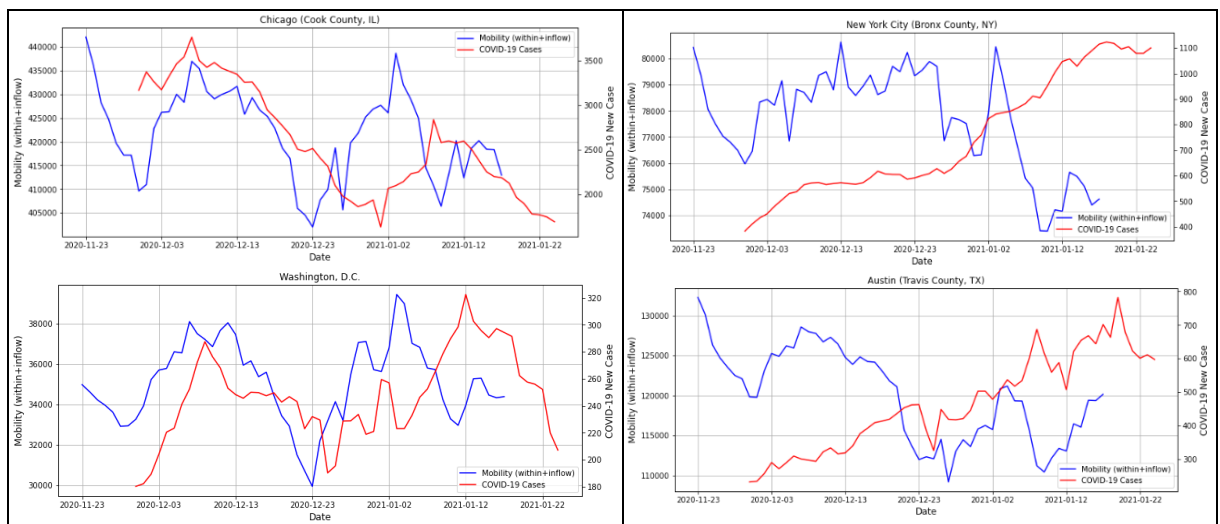

**Figure 2. Time-series graphs of four distinct counties. Two counties on the left have similar trends in mobility and COVID-19 case, while the other two on the right have dissimilar trends.**

## 5. Conclusion

Our study reveals that the association between human mobility and COVID-19 infections varies geographically and temporally in the U.S. through winter 2020. Our findings highlight time periods and locations (counties) where mobility may have influenced disease transmission. There are, however, some limitations in this study which we plan to address in our future work. First, we used a 7-day time lag between mobility and infections which is based on previous study findings on COVID-19 transmission. We plan to conduct a sensitivity analysis to evaluate the influence of the time lag from 0 to 30 days. Second, while DTW allows us to reveal similar trends of mobility and infections, it disregards negative associations between the series. Negative associations are also important to better understand human behavior because people tend to limit their movement when the number of infections increases. We plan to use differencing of the time series and apply other methods such as vector autoregression (Gao *et al.* 2009) and seasonal autoregressive integrated moving average model to capture negative associations (Williams *et al.* 1998). Third, we will compare pre-pandemic mobility data for the same season from the previous year as a control measure to evaluate the significance of our findings. Finally, external factors such as lockdown and social distancing policies and large gathering events such as protests may influence both human mobility and infections. We plan to identify and correlate these factors with the patterns we find in our results. This will allow us to identify the potential causes of the spatial heterogeneity and reveal if there are any confounding factors.